\documentclass[conference]{IEEEtran}

\usepackage{graphicx}

\usepackage{algorithm}
\usepackage{amsmath}
\usepackage{color}
\usepackage{amssymb}
\usepackage{soul}

\begin{document}

\newcommand{\mv}[1]{\ensuremath{\operatorname{\mathit{#1}}}}
\definecolor{dark}{gray}{.6}
\newcommand{\bc}[1]{\textcolor{dark}{#1}}
\newtheorem{lems}{Lemma}
\newtheorem{props}{Proposition}
\newtheorem{thms}{Theorem}
\newtheorem{defs}{Definition}
\newtheorem{obs}{Observation}

\title{Practical Experience Report: \\ The Performance of Paxos in the Cloud}

\author{
\IEEEauthorblockN{Parisa Jalili Marandi}
\IEEEauthorblockA{University of Lugano\\
Switzerland}
\and
\IEEEauthorblockN{Samuel Benz}
\IEEEauthorblockA{University of Lugano\\
Switzerland}
\and
\IEEEauthorblockN{Fernando Pedone}
\IEEEauthorblockA{University of Lugano\\
Switzerland}
\and
\IEEEauthorblockN{Ken Birman}
\IEEEauthorblockA{Cornell University\\
USA}
}

\maketitle

\begin{abstract}

This experience report presents the results of an extensive performance evaluation conducted using four open-source implementations of Paxos deployed in Amazon's EC2.
Paxos is a fundamental algorithm for building fault-tolerant services, at the core of state-machine replication.
Implementations of Paxos are currently used in many prototypes and production systems in both academia and industry.
Although all protocols surveyed in the paper implement Paxos, they are optimized in a number of different ways, resulting in very different behavior, as we show in the paper.
We have considered a variety of configurations and failure-free and faulty executions.
In addition to reporting our findings, we propose and assess additional optimizations to existing implementations.

\end{abstract}

\IEEEpeerreviewmaketitle


\section{Introduction}

In this experience report we present the results of an extensive performance evaluation we conducted with open-source implementations of Paxos~\cite{Lam98} deployed in Amazon's EC2, a public cloud-computing environment.\footnote{http://aws.amazon.com/ec2/}
Our study is motivated by the fact that many online services deployed in the cloud require both high availability and \emph{sustained performance}.
High availability is achieved by means of replication, using techniques such as state-machine replication~\cite{Sch90} and primary-backup replication~\cite{BMST93}.
At the core of these techniques lies an agreement protocol (e.g., \cite{Lam98, JunqueiraRS11}). A large variety of such agreement protocols exist in the literature that solve the problem under many different system assumptions~\cite{DUS04}. Among these protocols Paxos has received much attention in recent years both from the industry (e.g., \cite{MSR-TR-2005-33, CGR07}) and the academia (e.g., \cite{KA08, MRPTR2012, Ren11}).
Paxos has many desirable properties:
It is safe under asynchronous assumptions, live under weak synchronous assumptions, and resiliency-optimal, that is, it requires a majority of non-faulty processes (i.e., acceptors) to ensure progress.
Consequently, many distributed systems rely on Paxos for high availability
(e.g., Chubby~\cite{Bur06}, Spanner~\cite{Corbett:2012}, Autopilot~\cite{Isard07}).

Several recent papers have reported on the performance of Paxos implementations, mostly under ``normal conditions" (e.g., \cite{KA08, MRPTR2012,Marandi10,spaxos}), that is, deployments with homogenous nodes, balanced communication links, and the absence of failures.
While differences in the implementations impact overall performance, these reports typically show steady behavior in the normal case.
Yet, anecdotal evidence tells that under less favorable conditions (e.g., after the failure of a node), Paxos may lose its \emph{sustained performance}.  
Intuitively, this is explained by the fact that by relying on a quorum of acceptors for progress, Paxos may proceed at the pace of the quorum of faster acceptors, leaving slower acceptors lagging behind with an ever-increasing backlog of requests.
Paxos implementations generally read and process messages in arrival order, hence even if the messages in question relate to protocol actions that long since have been completed, they will be read and processed just as if they are associated with pending decisions. All of this will take time, hence should a fast acceptor fail and a slow one be needed to form a quorum, the system may experience a performance hiccup, corresponding to the time it takes for the slower acceptor to catch up.
%
Bursty behavior is undesirable because it can cascade into the application, within which end-user requests may be piling up and replicas falling behind. 

We set out to understand to what extent existing implementations genuinely suffer from this phenomenon and if so, under what conditions.
To this end we evaluated four open-source implementations of Paxos, S-Paxos, OpenReplica, Ring Paxos, and Libpaxos  under different message sizes in four configurations:
(a)~a homogeneous set of nodes in the same availability zone (i.e., datacenter);
(b)~two heterogeneous configurations with nodes in the same availability zone; and
(c)~homogeneous nodes distributed across different availability zones.
In each case we considered executions with and without participant failures.
These configurations represent the deployment of many current online services in the cloud.
Placing replicas on a set of nodes with similar hardware characteristics (configuration (a)) is probably the most common configuration used in experimental evaluations.
%
Heterogeneous settings (configuration (b)) may arise involuntarily (e.g., if applications run in a virtual machine whose physical node turns out to be shared among other applications) or voluntarily: a designer might choose to deploy Paxos in this manner, perhaps to reduce the perceived risk of correlated failures, or to reduce cost, for example by paying for 3 powerful nodes and 1 or 2 weaker backup nodes (e.g.,  
Cheap Paxos~\cite{LM04} is a variation of Paxos that exploits this alternative).
In addition to these two configurations, the participants of a service can be geographically distributed (configuration (c)) to improve locality and availability.
Locality reduces user-perceived latency and is achieved by moving the data closer to the users. Availability improves as the service can be configured to tolerate the crash of a few nodes within a datacenter or the crash of multiple sites. 

By evaluating the four open-source Paxos libraries under these configurations we show that standard Paxos implementations sometimes have unexpected behavior and long delays, although the phenomenon varies and depends on the details of the implementations: some protocols are more prone to problematic behavior; others are more robust but at the price of reduced performance. 
Our experience with the open-source libraries show that selecting the right implementation depends on the demands of the clients of the Paxos protocol and the environment in which a Paxos implementation will be deployed.

The remainder of this paper is organized as follows.
In Section~\ref{sec:bckg} we briefly review Paxos.
In Section~\ref{libs} we describe the libraries used in our performance evaluation with emphasis on their flow control mechanisms, and propose optimizations to one of them. 
In Section~\ref{sec:eval} we detail our experimental setup and present the results. In Section~\ref{sec:lessons} we discuss the main lessons we have learnt while interacting with these libraries and we conclude the paper in Section~\ref{sec:final}.

\section{Background}
\label{sec:bckg}

Paxos assumes a distributed system model in which processes communicate by exchanging messages. 
Processes may fail by crashing but never perform incorrect actions.
Although there are variations of Paxos that tolerate arbitrary process behavior (e.g.,~\cite{Castro:2002, RamasamyC05}), such failures are not the focus of this paper.

Paxos distinguishes the roles of \emph{proposer}, \emph{acceptor}, and \emph{learner}, where a process can play one or more roles simultaneously.
Each instance of Paxos proceeds in two phases:
In Phase 1, the \emph{leader} or \emph{coordinator} (e.g., a process among the acceptors or proposers) selects a unique round number \emph{c-rnd} and asks the acceptors to promise that in the given instance they will reject any requests (Phase 1 or 2) with round number less than \emph{c-rnd}.
Phase 1 is completed when a majority-quorum $Q_a$ of acceptors confirms the promise to the leader.
Notice that since Phase 1 is independent of the value proposed it can be pre-executed by the leader~\cite{Lam98}.
If any acceptor already accepted a value for the current instance, it will return this value to the leader, together with the round number received when the value was accepted (\emph{v-rnd}).

Once a leader completes Phase 1 successfully, it can proceed to Phase 2. 
In Phase 2, the leader selects a value according to the following rule: if no acceptor in $Q_a$ accepted a value, the leader can select any value. 
If however any of the acceptors returned a value in Phase 1, the leader is forced to execute Phase 2 with the value that has the highest round number \emph{v-rnd} associated to it. 
In Phase 2 the leader sends a message containing a round number (the same used in Phase 1). 
Upon receiving such a request, the acceptors acknowledge it, unless they have already acknowledged another message (Phase 1 or 2) with a higher round number. 
Acceptors update their \emph{c-rnd} and \emph{v-rnd} variables with the round number in the message. 
When a quorum of acceptors accepts the same round number (Phase 2 acknowledgement), consensus terminates: the value is permanently bound to the instance, and nothing will change this decision. 
Thus, learners can deliver the value. 
Learners learn this decision either by monitoring the acceptors or by receiving a decision message from the leader.

As long as a nonfaulty leader is eventually selected and there is a majority quorum of nonfaulty acceptors and at least one nonfaulty proposer, every consensus instance will eventually decide on a value.
A failed leader is detected by the other nodes, which select a new leader. 
If the leader does not receive a response to its Phase 1 message it can re-send it, possibly with a bigger round number.
The same is true for Phase 2, although if the leader wants to execute Phase 2 with a higher round number, it has to complete Phase 1 with that round number.
Eventually the leader will receive a response or will suspect the failure of an acceptor.


%
%
%
%
\section{Open-source Paxos libraries}
\label{libs}

In our evaluation, we worked with four open-source Paxos implementations.
Recall that Paxos requires a majority-quorum for progress (i.e., it remains operational despite the failure of $f$ acceptors out of 2$f$+1).
As soon as a participant (e.g., leader) receives a majority of Phase 2B messages for a value in an instance, the participant knows the instance is decided.
We call this quorum the participant's \emph{first majority-quorum}.
Different participants may have distinct first majority-quorums, but if an acceptor is ``slow", then it is unlikely to participate in any first majority-quorum.
In fact, one can expect that a first majority-quorum will likely contain ``fast" acceptors only. 

An acceptor can be slow for many reasons.
For example, perhaps the slow acceptor cannot keep up with the fast acceptors because it is running on a node with less processing power than the fast acceptors or its CPU is shared among several processes.
It could also be that its communication links are subject to higher latencies than the other nodes' links.
Whatever the reason, the notions of slow and fast acceptors are important because Paxos is quorum-based, moving from one instance to the other as soon as a majority of acceptors is prepared to do so.
In the following, we argue that in principle such a distinction between acceptors may have performance implications, notably in the case of failures.
In the subsequent section, we assess this phenomenon experimentally.

\subsection{S-Paxos}
S-Paxos~\cite{spaxos} is implemented in Java.\footnote{https://github.com/nfsantos/S-Paxos}
It is composed of a set of replicas, each one playing the combined roles of proposer, acceptor and learner.
One of the replicas is elected the leader. 
The key idea in S-Paxos is to load-balance request reception and dissemination among all the replicas. 
A client selects an arbitrary replica and submits its requests to it. 
After receiving a request, a replica forwards it (or possibly a batch of requests) to all the other replicas. 
A replica that receives a forwarded request sends an acknowledgement to all the other replicas. 
When a replica receives $f+1$ acknowledgements, it declares the request stable.
This is needed because in S-Paxos ordering is performed on request ids.
As in classic Paxos, the leader is responsible for ordering requests. 
A participant considers an instance decided after receiving $f+1$ Phase~2B messages from the acceptors. 
All the replicas execute all the requests but only the replica who receives the request responds to the client. 
S-Paxos strives to balance CPU and network resources, but many messages must be exchanged before a request can be ordered. 
Due to the high number of messages exchanged, S-Paxos is CPU-intensive and benefits from deployment on powerful multi-core machines.  

S-Paxos uses blocking I/O for the communications among replicas. 
As mentioned earlier, a replica forwards batches of requests to all the other replicas. 
If a replica is slow in handling its incoming traffic, another replica will block upon sending new messages to the slow replica since communication is based on TCP.
Thus, faster acceptors cannot transfer more batches to the slow replica and we expect the performance of the system to follow the speed of the slowest replica. 
%
Moreover, since S-Paxos is designed around the idea of distributing the load among acceptors, reducing the number of acceptors (e.g., due to failures) may result in reduced performance.

\subsection{OpenReplica}

OpenReplica is an open-source library implemented in Python\footnote{https://pypi.python.org/pypi/concoord} that enables automatic replication of user-provided objects~\cite{Altinbuken12}. 
OpenReplica is composed of a set of replicas and a set of acceptors. 
Replicas are the processes that replicate an object; in Paxos's parlance, they play the ``learner" role. 
One of the replicas is also the leader in Paxos. In OpenReplica clients send their requests to a client proxy who batches the requests. The client proxy then connects to the pool of replicas to send the batched requests and OpenReplica ensures that the requests are forwarded to the leader to be ordered. Replicas deliver and execute the sequence of requests in the order dictated by instance identifiers. After executing a request replicas respond to the clients. 

The leader in OpenReplica uses non-blocking I/O to communicate with the acceptors. 
If the transmission of a message to an acceptor cannot happen immediately (e.g., because the communication buffer associated with the acceptor is full), the leader is notified and retries the transmission until it succeeds. 
If an acceptor is slower than the others, its buffers will fill up faster and communications with it will cause retransmissions at the leader. This affects performance because the leader will work harder and some portion of its I/O bandwidth will be lost to retransmissions. If during the time it takes for the slow acceptor to catch up a fast acceptor crashes, we can expect a further reduction in performance since a majority-quorum of acceptors will not be available immediately given that the slow acceptor is needed to form a majority-quorum. 

\subsection{Ring Paxos}
\label{subsec:ring-paxos}

Ring Paxos is an open-source Paxos library implemented in Java.\footnote{https://github.com/sambenz/URingPaxos}
(There is also a C implementation of Ring Paxos\footnote{http://sourceforge.net/projects/libpaxos/files/RingPaxos/} that relies on ip-multicast; we use the Java version, based entirely on unicast communication.) 
Ring Paxos disseminates processes on a logical uni-directional ring to make a balanced usage of the available bandwidth~\cite{marandi2014ring}. 
A process in Ring Paxos can play the roles of the acceptor, proposer, learner, and coordinator. 
One of the acceptors is elected as the leader.
Ring Paxos handles leader election and the ring's configuration via Zookeeper.\footnote{http://zookeeper.apache.org/} 
Clients submit their requests to those processes in the ring that assume the role of proposers. 
Proposers batch the requests and forward them along the ring. 
The leader initiates Paxos for the batches of requests that it assembles and the batches it receives from other processes in the ring. 
Acceptors create Phase~2B messages and send them to their successors. 
Processes that are not acceptors simply forward Phase~2B messages they receive to their successors.
The final decision is made by the acceptor that receives $f+1$ Phase~2B messages. 
The decision circulates in the ring until all processes receive it. 
The learners deliver instances following instance identifiers. 



Processes in the ring communicate using TCP; learners send replies to the clients through UDP. 
All communications is based on non-blocking I/O. 
Both clients and processes in the ring can batch messages.
In a ring, a slow process can negatively affect the overall performance as it may become a system's bottleneck. 
We expect the ring to operate at the speed of the slowest acceptor. 
If an acceptor leaves the ring, Ring Paxos will reconfigure the ring and during reconfiguration performance may suffer. 

\subsection{Libpaxos} 

Libpaxos is implemented in C.\footnote{https://bitbucket.org/sciascid/libpaxos}
It distinguishes proposers, acceptors, and learners, where proposers are also learners. Libpaxos does not handle leader election. 
Applications must decide how to ensure the existence of a single leader (e.g., one option is to use Zookeeper).
To submit requests, clients connect directly to the proposers. 
Acceptors send their Phase~2B messages, including the agreed value, to the proposers and to the learners. 
Upon receiving $f+1$ Phase~2B messages from the acceptors, the learners and the proposers declare an instance as decided. 
The learners deliver instances following instance identifiers. 

Processes in Libpaxos communicate using non-blocking buffered I/O provided by the libevent library.\footnote{http://libevent.org} Libpaxos does not explicitly batch the requests; batching is implemented by the buffered communication provided by libevent. 
Besides sending Phase~2B messages, an acceptor also sends values to the learners and proposers, therefore, the acceptor's outgoing traffic is higher than its incoming traffic.
A slow acceptor may become overwhelmed by a high volume of incoming messages, in which case messages will pile up at the sender's side, or by a high rate of outgoing messages, in which case messages will pile up at acceptor's side.

In either case, until a slow acceptor becomes overwhelmed, performance in Libpaxos will be driven by the faster acceptors.
If a fast acceptor crashes and a slow acceptor is needed to form a majority quorum, the system may experience periods of inactivity until the slow acceptor processes its backlog of requests.


\begin{table*}[ht]

%
\centering
\small

\begin{tabular}{|c|c|c|c|c|c|c|c|} \hline
Configuration &  Type & Environment &  Leader$^\ast$ & ~~A1~~ & ~~A2~~ & ~~A3~~ & Learner$^\dagger$ \\  \hline \hline

(a) & Homogeneous & LAN & \multicolumn{5}{c|}{Small}  \\ \hline

(b) & Heterogeneous & LAN  & \multicolumn{3}{c|}{Small} & Micro & Small\\ \hline

(c) & Heterogeneous & LAN &Large & \multicolumn{2}{c|}{Small} & Micro & Small\\ \hline

(d) & Homogeneous$^\ddagger$ & WAN &\multicolumn{5}{c|}{Small}  \\ \hline
\end{tabular}
\vspace{5mm}
\caption{Configurations used in the evaluations.
(Legend: $^\ast$The leader in S-Paxos and Ring Paxos is also acceptor A1 and the leader in OpenReplica is also the replica. $^\dagger$The concept of an independent learner only exists in Libpaxos and Ring Paxos. $^\ddagger$ Although machines in this configuration are homogenous, the acceptor in the remote datacenter, A3, is connected to other processes with a lower bandwidth.)}
\vspace{-7mm}
\label{table:configs}
%
\end{table*}

\subsection{Libpaxos$^+$}
\label{libplus}

Motivated by our observations on the behavior of Libpaxos, presented in the next section, we created Libpaxos{$^+$}, an extension to the original protocol.
The key idea is for proposers to selectively involve acceptors in Paxos's Phase 2 based on how the acceptors performed in previous instances.
If an acceptor was not in the first majority-quorum of past instances, then it might be a slow acceptor and should be spared in the next few instances.
Libpaxos{$^+$} thus attempts to reduce the backlog of slow acceptors in order to allow them to catch up, so that they can achieve better response times later in instances in which they participate.


We modify a proposer so that its execution is divided into \emph{steps}, where a step is a sequence of Paxos instances.
In the first instances in the step, the proposer sends Phase 2A messages to all acceptors and records the number of instances each acceptor is included in the first majority-quorum.
In the following instances in the step, the proposer sends Phase 2A messages to a majority-quorum only, composed of those acceptors who appeared most often in the initial instances.
A step finishes when a pre-determined number of instances are executed or the proposer suspects the crash of an acceptor among the selected ones to participate in the instance.



\begin{figure*}[ht]

\vspace{-4mm}
\center
    \begin{tabular}{c}
         \includegraphics[width=0.85\textwidth]{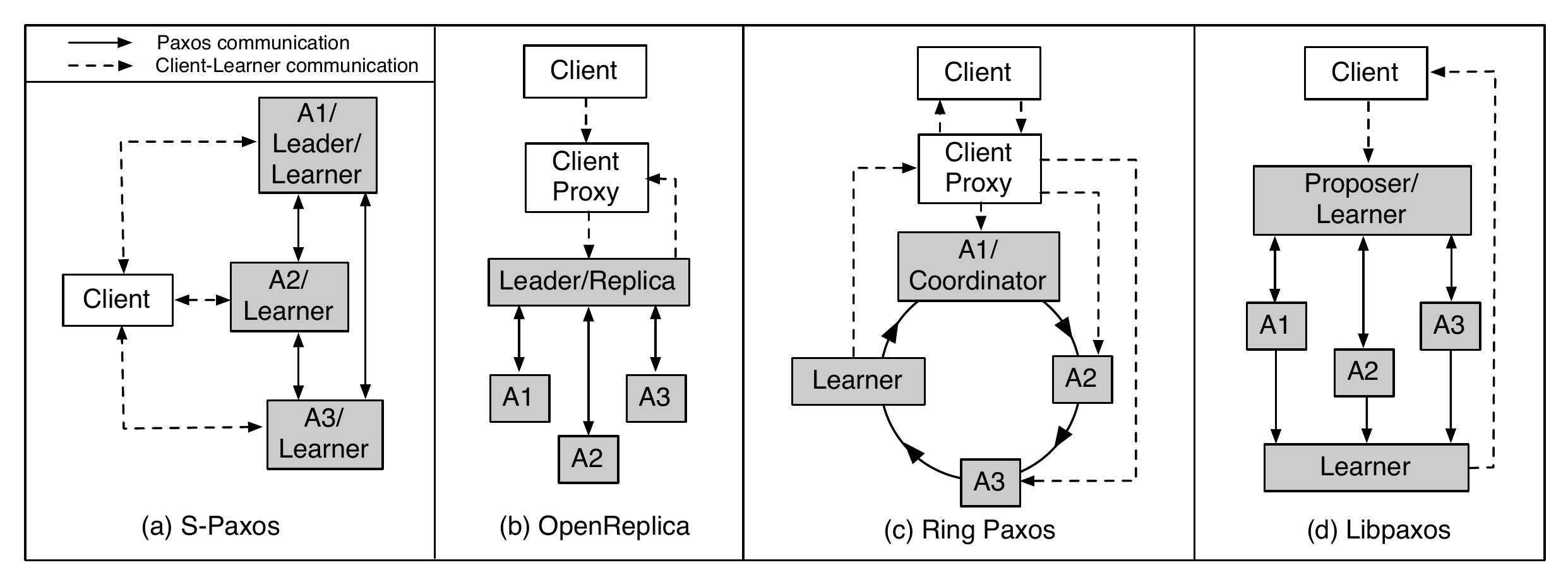}
     \end{tabular} 
        \caption{The communication pattern and architectural differences among the four libraries as deployed in the experiments; $f$ is equal to one; the picture preserves the terminology of the libraries; dashed lines show the communications between the learner/replica and the clients; in all the libraries the learner/replica sends the responds back to the client; in S-Paxos and OpenReplica clients send their requests to the nodes that also assume the learner/replica role.}
    \label{fig:architecture}
    \vspace{-4mm}
\end{figure*}

\section{Experimental evaluation}
\label{sec:eval}

In this section, we describe the experimental setup, explain our methodology for the experiments, report on the peak performance of each library under various conditions, and analyze each library under failures.

\renewcommand{\arraystretch}{1.5}

\subsection{Experimental setup}

\noindent \textbf{Hardware setup.} All the experiments are performed in Amazon's EC2 infrastructure with a mix of small, micro, and large instances, as detailed next. In all the experiments each process runs on a separate Amazon EC2 instance. In the following, one EC2 compute unit provides the equivalent CPU capacity of a 1.0-1.2 GHz 2007 Opteron or 2007 Xeon processor, vCPU represents the number of virtual CPUs for the instance.
\begin{itemize}
\item Micro: up to 2 ECUs (EC2 Compute Unit), 1 vCPUs, 0.613 GBytes memory, very low network capacity.
\item Small: 1 ECUs, 1 vCPUs, 1.7 GBytes memory, 1x 160 GBytes of storage, low network capacity.
\item Large: 4 ECUs, 2 vCPUs, 7.5 GBytes memory, 2x 420 GBytes of storage, moderate network capacity. 
\end{itemize}

All servers run Ubuntu Server 12.04.2-64 bit and the socket buffer sizes are equal to 16 MBytes. 

\noindent \textbf{Configurations.} We measure the performance of S-Paxos, OpenReplica, Ring Paxos, and Libpaxos in four different configurations (see Table~\ref{table:configs}). In S-Paxos and Ring Paxos one of the acceptors plays the role of the leader and thus in Table~\ref{table:configs} the leader represents acceptor A1 for S-Paxos and Ring Paxos. For each configuration, all the libraries are evaluated with three request sizes: 200 Bytes, 4 KBytes, and 100 KBytes. All the libraries are in-memory in our experiments. Ring Paxos relies on Zookeeper for ring's configuration. Session timeout for the Zookeeper is set to 3 seconds in all the experiments. 

In the experiments performed in a LAN, all the instances are deployed in the US-West-2c region. In the experiments performed in a WAN, processes are distributed among three availability zones: In the experiments with Libpaxos and OpenReplica the leader, A2, the learner, and clients are located in US-West-2c, A1 is located in US-West-2b, and A3 is located in US-East-1b. In the experiments with S-Paxos and Ring~Paxos the leader (A1), the learner (in Ring Paxos), and the clients are located in US-West-2c, A2 is located in US-West-2b, and A3 is located in US-East-1b. moreover, in all the experiments with Ring Paxos, a stand alone version of Zookeeper is deployed on a micro instance located in US-East-1c. 
As a reference, the RTT value is 1.5 ms (millisecond) in US-West-2c, 3.9 ms between US-West-2b and US-West-2c, 82 ms between US-West-2c and US-East-1b, and 90 ms between US-West-2b and US-East-1b.

\begin{figure*}[ht]
\vspace{-20mm}
    \begin{tabular}{c}
 	 \includegraphics[width=\textwidth]{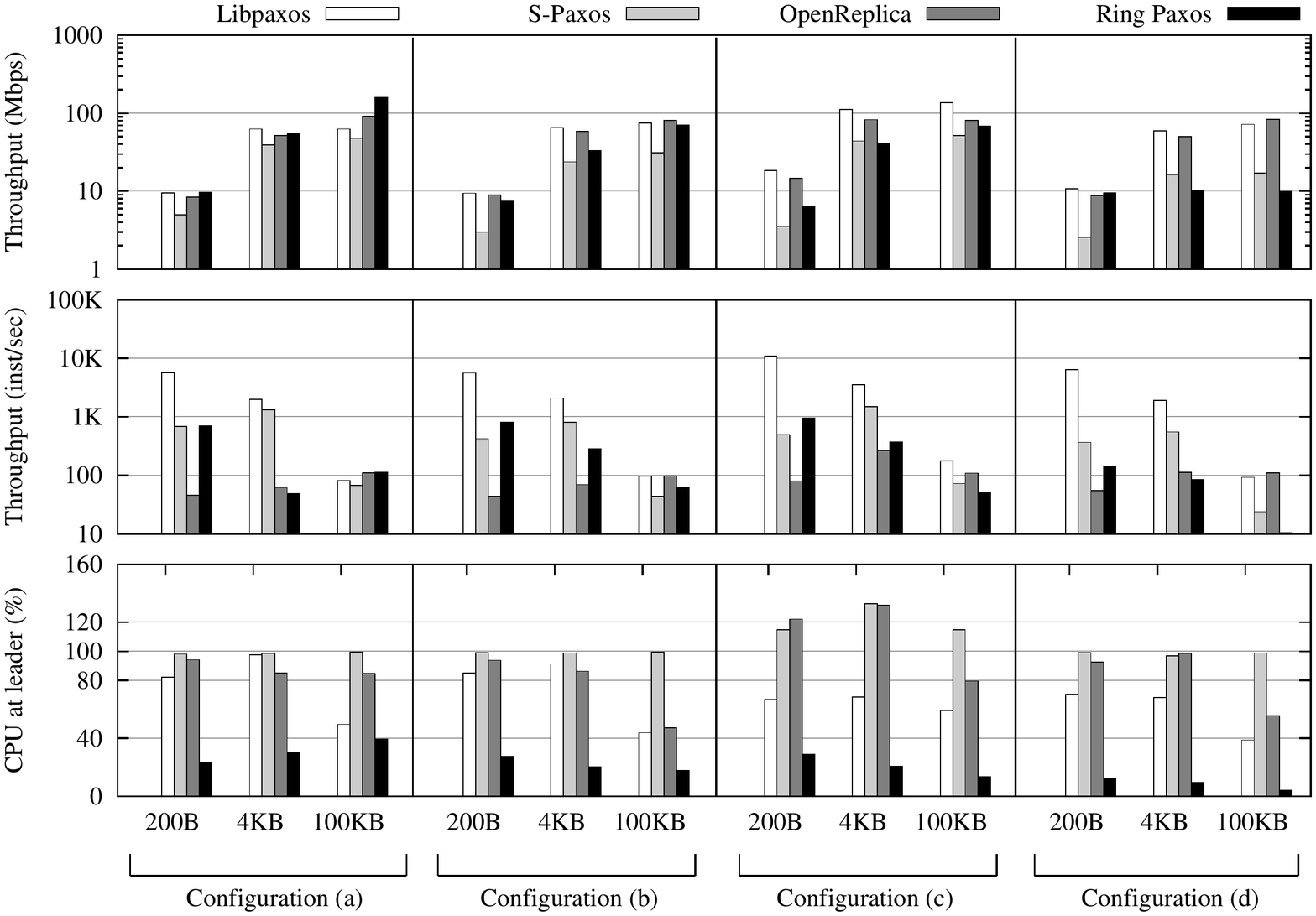}
     \end{tabular} 
	\vspace{-7mm}
        \caption{Peak performance of Libpaxos, S-Paxos, Ring Paxos, and OpenReplica in four configurations (see Table~\ref{table:configs}); y-axis in the two top-most graphs is in log scale; note that S-Paxos, Ring Paxos, and OpenReplica are multithreaded and therefore in some scenarios (configuration (c)) the CPU usage at the leader is higher than 100\% for some of the libraries.}
    \label{fig:all-libraries}
         \vspace{-3mm}
\end{figure*}

\noindent \textbf{Architectural differences.} Figure~\ref{fig:architecture} illustrates the inter-process communication patterns and architectural differences of the four libraries while preserving their specific terminology (for the details see Section~\ref{libs}). Notice that the terms \emph{leader}, \emph{proposer}, and \emph{coordinator} convey the same concept and so do the terms \emph{replica} and \emph{learner}. In Ring Paxos, however, proposer refers to any node that receives requests from clients and forwards them to other processes. In all the experiments, there are three acceptors in all the libraries and $f$ is equal to one. In S-Paxos and Ring Paxos, the leader (or the coordinator) role is assumed by one of the acceptors (acceptor A1). In OpenReplica and Libpaxos, however, a separate process is elected as the leader (or proposer).\footnote{We emphasize that the differences in the deployments are due to the unique properties of the libraries rather than the choices made by the authors.} 

In the experiments with OpenReplica, each client process has a client proxy (as a separate module in the client process) that batches the requests of the client and sends them to the leader. The client waits for the responses before submitting new requests. Similarly to OpenReplica in Ring Paxos a client has a module for batching the requests. The client proxy batches the requests and sends them to an acceptor that also plays the role of proposer. The size of a batch is 12 KBytes in all the experiments. Batching in Ring Paxos has been disabled throughout the experiments. In the experiments with S-Paxos, a client sends a request to a randomly chosen replica and waits for its response before sending a new request. A replica batches requests before disseminating them to the other replicas. In our experiments this batch size is 1KByte. 
Also note that the batch sizes in these two libraries are chosen to get the best performance. In the experiments with Libpaxos, to ensure progress we have configured a single proposer. Clients send a request to the proposer and wait for the request's response from the learner before sending a new request. 

\subsection{Methodology}
\label{methodlogy}

The goal of our performance assessment is twofold: first, we measure the peak performance of S-Paxos, OpenReplica, Ring Paxos, and Libpaxos in a set of configurations as illustrated in Table~\ref{table:configs} (see Section~\ref{pperf} for details). Second, we select a subset of these configurations to take a closer look at the flow control mechanisms of the libraries (see Sections~\ref{subsec:spaxos},~\ref{subsec:openreplica},~\ref{subsec:ringpaxos},~\ref{subsec:libpaxos}). Since libraries are different in their implementation and communication strategies, the configurations we choose vary across libraries. Table~\ref{table:failconfigs} enlists the set of the chosen configurations. 
We note that space limitations and the large variety of possible configurations force us to focus on a few cases.
These results were selected from a much larger set of experiments we have performed and were chosen to clearly reveal each library's behavior.

\begin{table}[ht]
\centering
\small
\begin{tabular}{|l|l|} \hline
Library &  [Conf. ,Size]\\  \hline \hline
S-Paxos &[(b),4] [(d),4]\\ \hline
OpenReplica&[(c),100] [(d),4] \\ \hline
Ring Paxos&[(b),100] [(d),4]  \\ \hline
Libpaxos&[(b),4]  [(d),4]\\ \hline
\end{tabular}
\vspace{5mm}
\caption{Configurations in which we evaluate the flow control mechanism of the open-source libraries.}
\vspace{-5mm}
\label{table:failconfigs}
\end{table}

\subsection{Peak performance}
\label{pperf}

\begin{figure*}[ht]
    \begin{tabular}{c@{}c}
\includegraphics[width=\columnwidth]{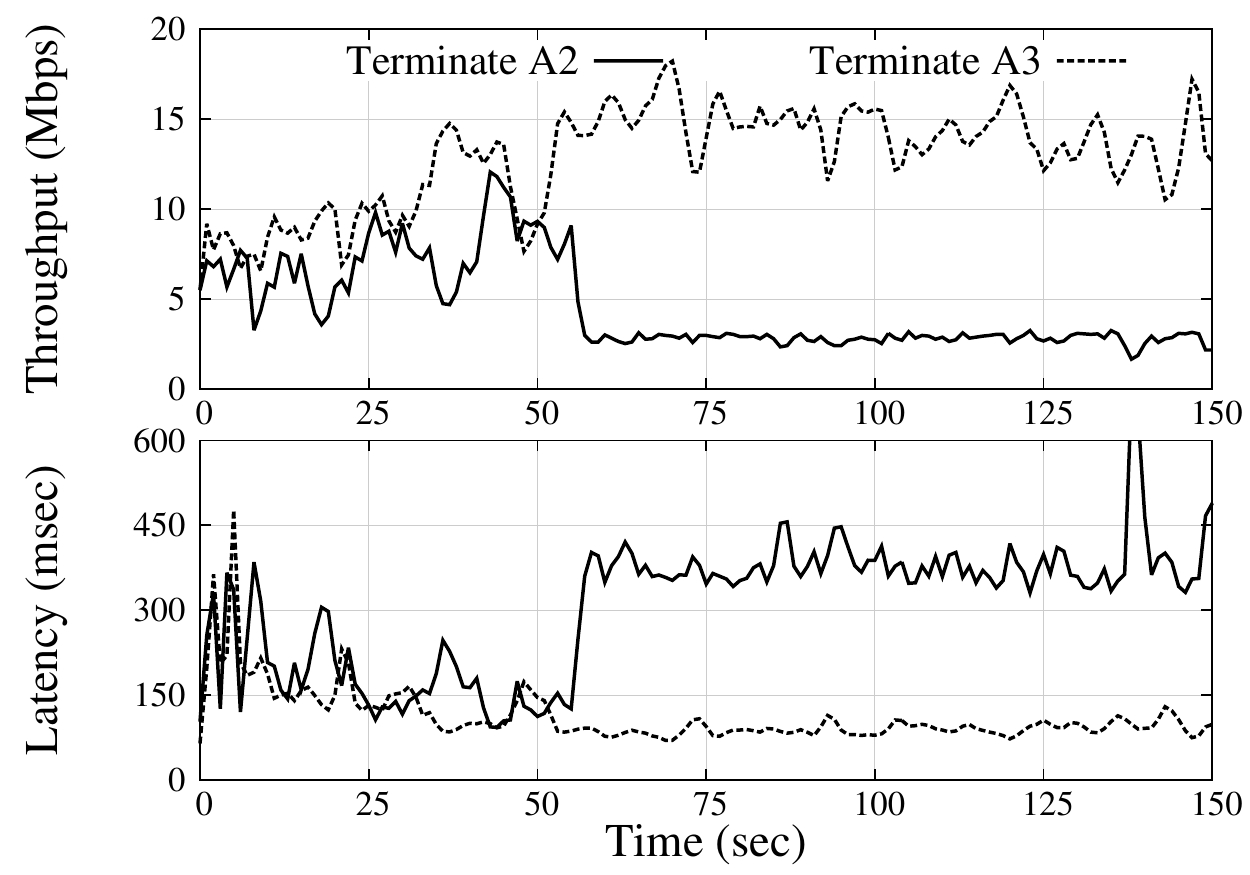} &
\includegraphics[width=\columnwidth]{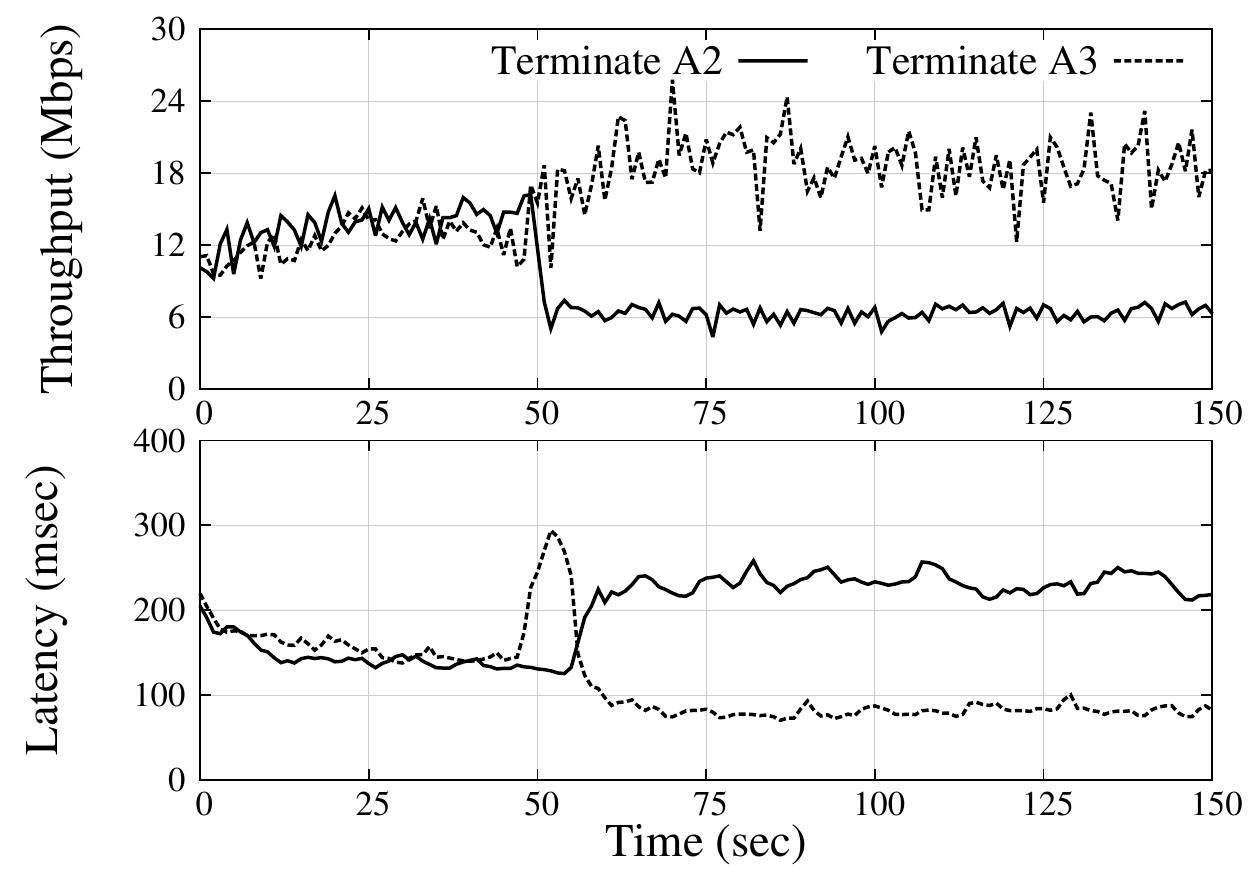} \\
    \end{tabular}
	\caption{Performance of \textbf{S-Paxos} in configuration (b) with 4 KByte requests at 70\% of peak throughput (left-most graphs); and in configuration (d) with 4 KByte requests at peak performance two experiments are performed (right-most graphs); in each configuration two experiments are performed; at each experiment one acceptor (A2 or A3) is terminated after 50 seconds.}
    \label{fig:s-paxos}
\end{figure*}
\begin{figure*}[ht]
    \begin{tabular}{c@{}c}
\includegraphics[width=\columnwidth]{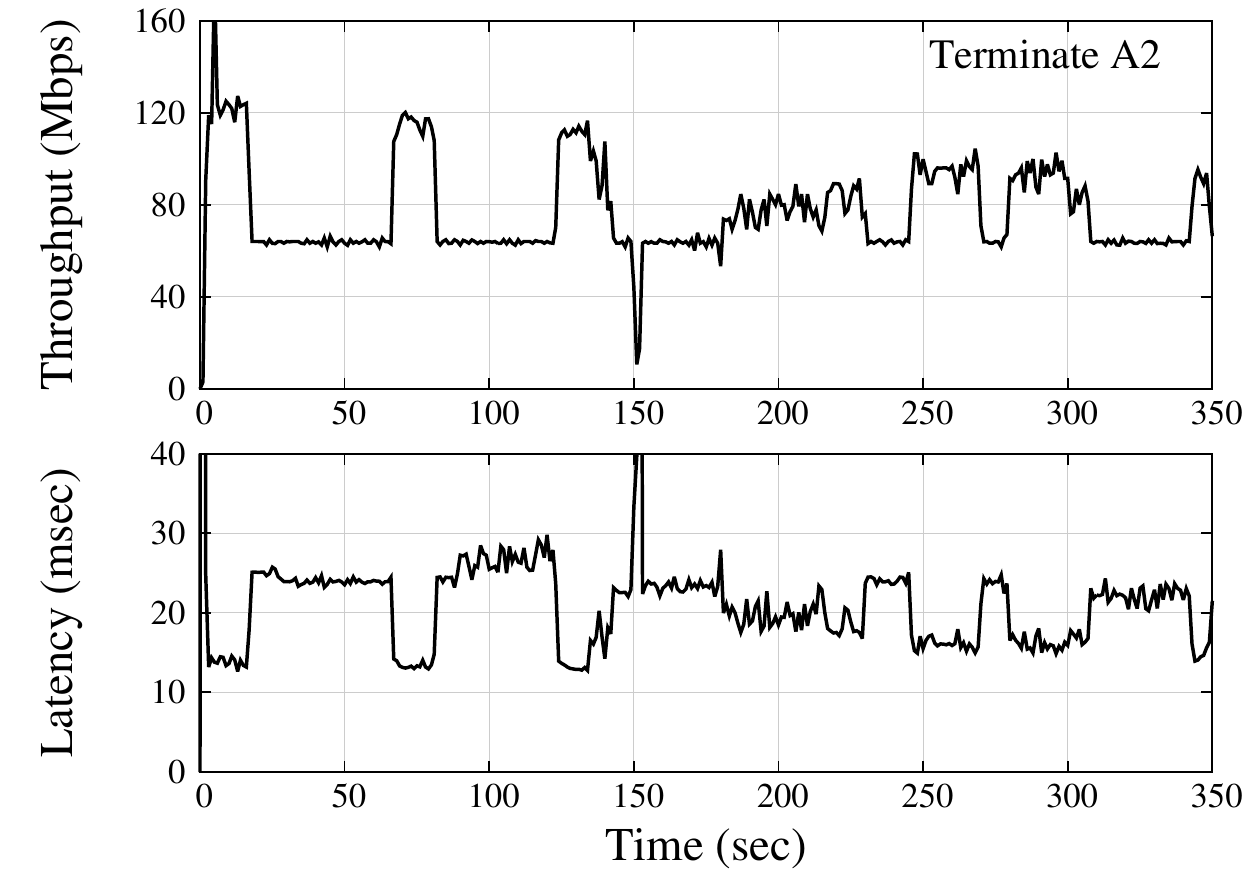} &
\includegraphics[width=\columnwidth]{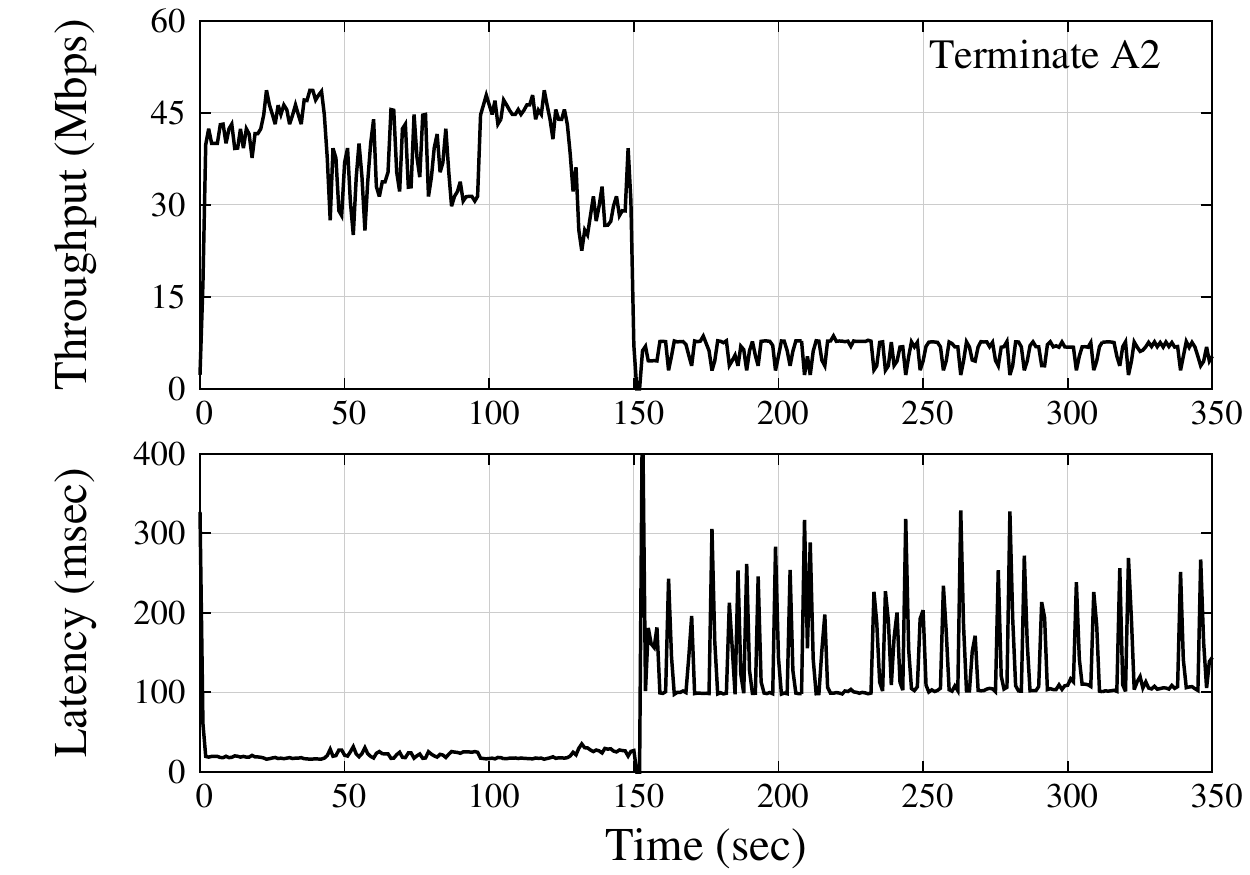} \\
    \end{tabular}
\caption{Performance of \textbf{OpenReplica} in configuration (c) with 100 KByte requests at peak performance (left-most graphs);  and in configuration (d) with 4 KByte requests at 70 \% of peak performance (right-most graphs); In both the configurations acceptor A2 is terminated after 150 seconds.}
\vspace{-5mm}
    \label{fig:openreplica}
\end{figure*}
\begin{figure*}[ht]
    \begin{tabular}{c@{}c}
\includegraphics[width=\columnwidth]{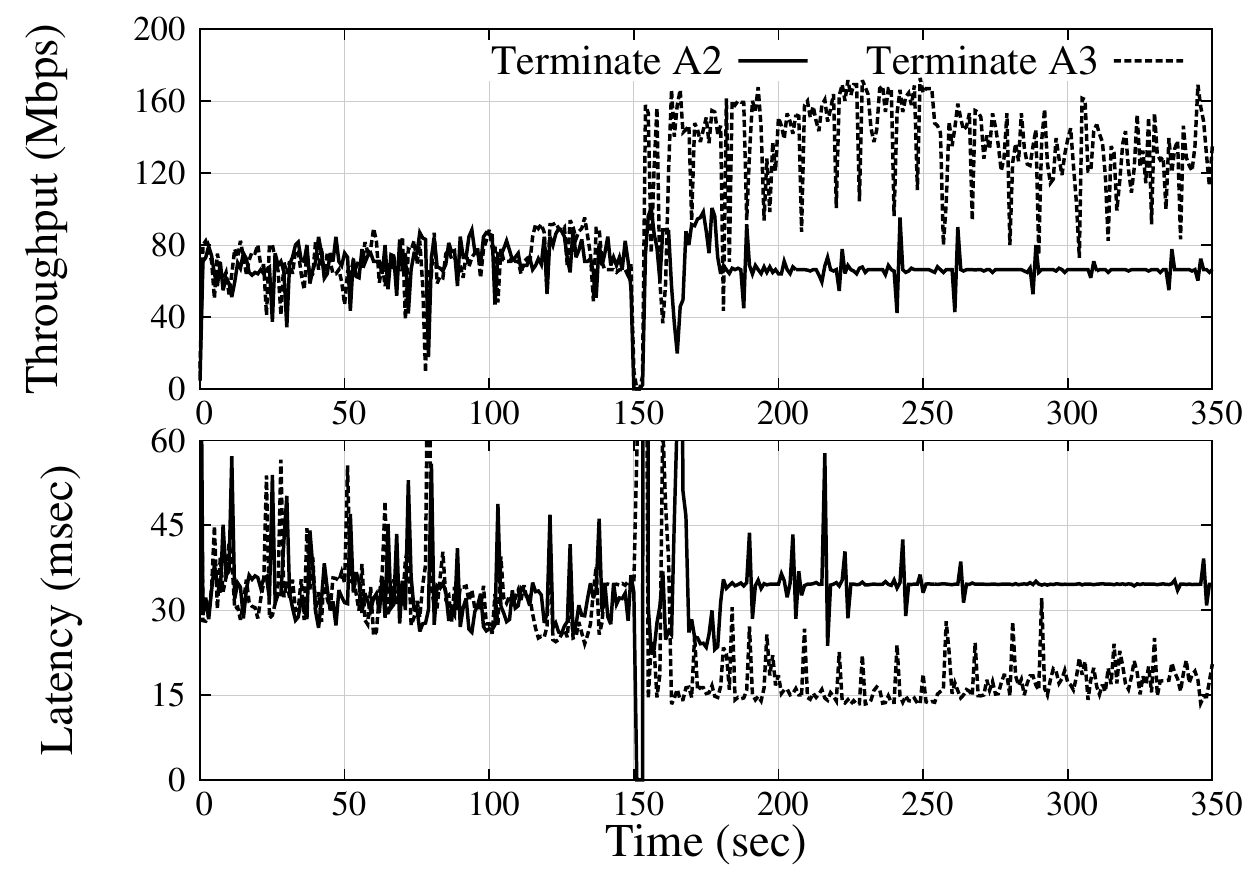}&
\includegraphics[width=\columnwidth]{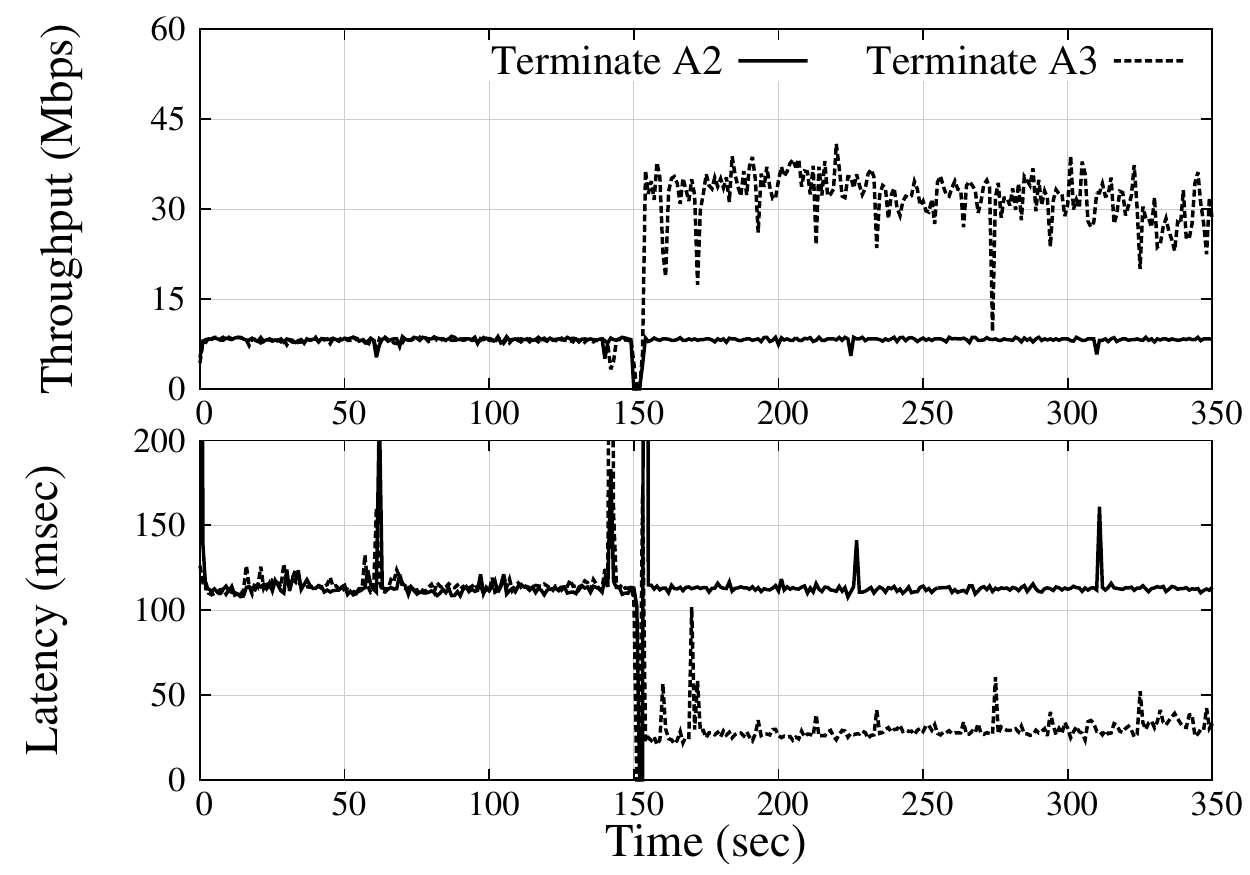} \\
    \end{tabular}
\caption{Performance of \textbf{Ring Paxos} in configuration (b) with 100 KByte requests at 100\% of the peak performance (left-most) graphs;  and in configuration (d) with 4 KByte requests at 70\% of the peak performance (right-most graphs); in each configuration two experiments are performed; in each experiment one acceptor (A2 or A3) is terminated after 150 seconds.}
\vspace{-4mm}
    \label{fig:ringpaxos}
\end{figure*}

Figure~\ref{fig:all-libraries} displays the results for peak performance. 
The graphs in this figure measure the following performance metrics from top to bottom: delivery throughput in megabits per second, delivery throughput in number of decided instances per second, and CPU usage at the leader. Note that the number of requests delivered per second is directly proportional to the delivery throughput in megabits per second. Each experiment is performed for a period of 100 seconds and the first and last 10 seconds are discarded. S-Paxos, Ring Paxos, and OpenReplica are multithreaded and therefore in some scenarios (configuration (c)) the CPU usage at the leader is higher than 100\%. 

The following patterns can be discerned from Figure~\ref{fig:all-libraries}. When comparing the throughput, unless stated otherwise, the values of throughput in Mbps are considered (top-most graph). 

\begin{itemize}
\item For all implementations and configurations, as we would expect, throughput improves as the request size increases, although the improvement is more noticeable from small to medium messages. In most of the configurations, OpenReplica and Libpaxos show similar performance, better than S-Paxos and Ring Paxos's performance. 

 \item We now assess each protocol when we replace one of the small acceptors in configuration (a) with a slower acceptor (see configuration (b) in Table~\ref{table:configs}). Regardless of request size, performance of Libpaxos and OpenReplica does not change between configurations (a) and (b) since their execution is driven by the fastest majority-quorum and in both configurations there is a majority-quorum that contains two small acceptors. The throughput of S-Paxos and Ring Paxos in configuration (b) is lower than in configuration (a) since S-Paxos and Ring Paxos adapt their performance to the speed of the slowest member. 

\item When we placed the leader in configuration (b) in a more powerful node, configuration (c), we observed that in all protocols except Ring Paxos the throughput increased, regardless of the request size. The leader of Ring Paxos is not CPU-bound in configuration (b) and therefore replacing the leader by a stronger machine had no effect in its performance. 

\item To understand the effects of geographical deployments on the performance, compare the results in configurations (a) and (d). The performance of Libpaxos and OpenReplica do not change between the two configurations: in both configurations there is a majority-quorum in the vicinity of the proposer (leader) in the two libraries. Hence, Libpaxos and OpenReplica are not limited by the slow links between the proposer and the acceptor located in a remote region (US-East). Performance of S-Paxos and Ring Paxos on the other hand is dictated by the slowest link. 

\item In most of the configurations and with small requests the leader in Libpaxos and OpenReplica is CPU-bound. Except for configuration (c), S-Paxos is always CPU-intensive. This happens because threads constantly spin while waiting for events (e.g., a message to arrive). Ring Paxos is never CPU-bound. 

\item Although all the implementations achieve more or less comparable peak throughput (in Mbps), the number of instances decided per second varies across them. We attribute this to differences in their batch sizes and also the processes in which batching takes place. 
\end{itemize}

\subsection{S-Paxos under failures} 
\label{subsec:spaxos}
Figure~\ref{fig:s-paxos} shows the performance of S-Paxos in configuration (b) with 4 KByte requests, under 70\% of peak throughput and also in configuration (d) with 4 KByte requests over a period of 150 seconds and at peak throughput. The top graphs show the delivery throughput in megabits per second and the bottom graphs show the corresponding latency in milliseconds. In the experiments, after 50 seconds of the execution one acceptor is terminated. 
In S-Paxos the load is distributed among acceptors and the execution proceeds at the pace of the slowest or the most distant acceptor. Thus, in both configurations after the termination of acceptor A3, throughput increases and latency decreases. This happens because performance is no longer limited by the slow acceptor. However, after the termination of acceptor A2, throughput decreases and latency increases. This is a consequence of the fact that acceptor A2 no longer contributes its share to the performance.

\begin{figure*}[ht]
    \begin{tabular}{c@{}c}
     \includegraphics[width=\columnwidth]{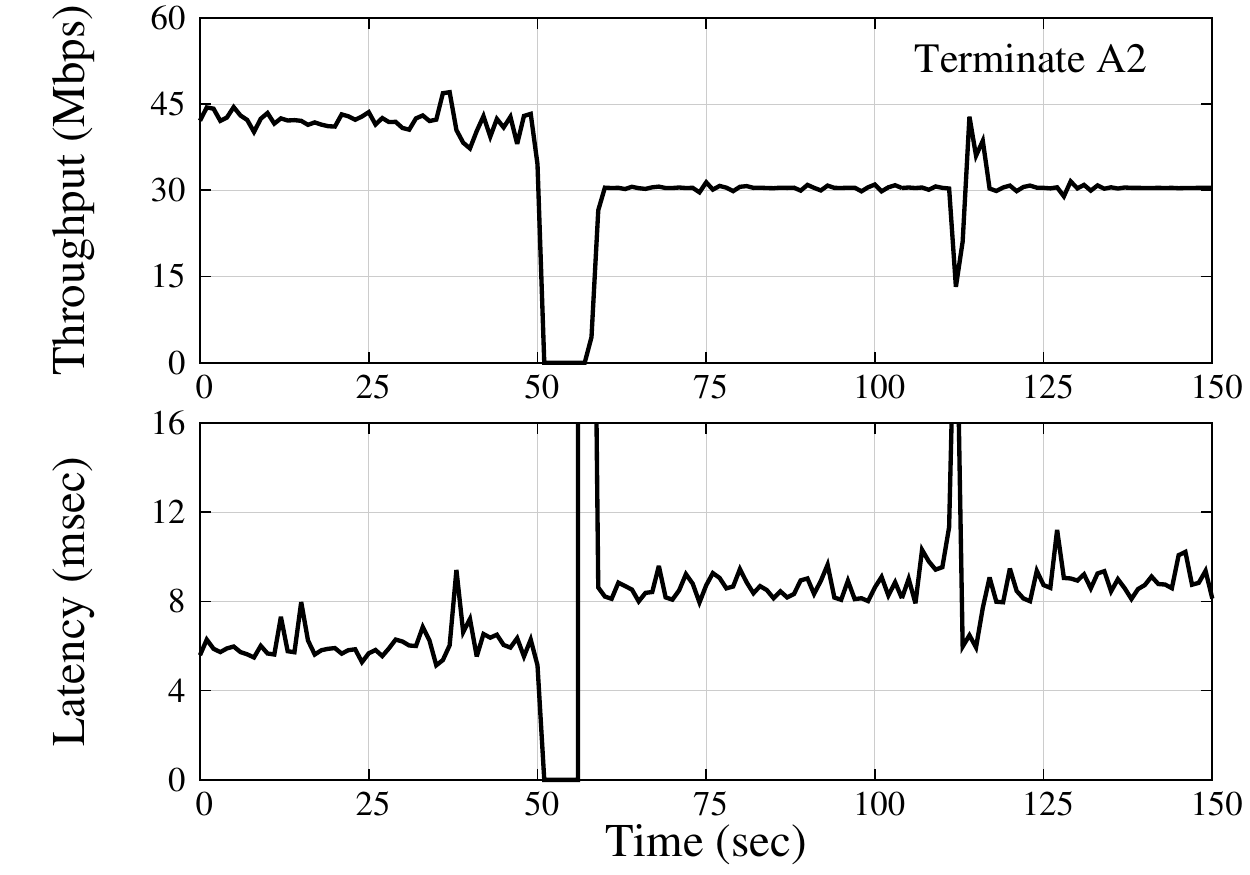}&
      \includegraphics[width=\columnwidth]{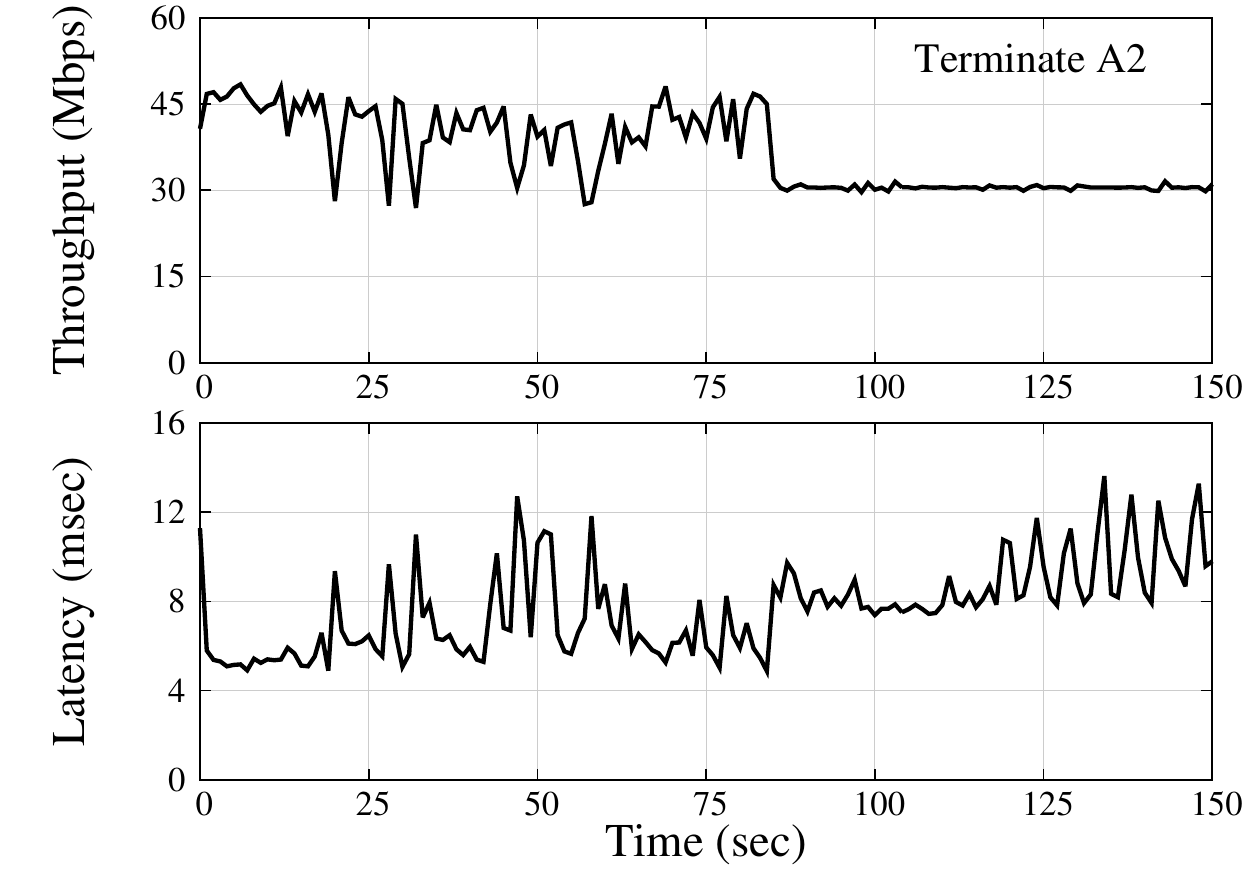} \\
           \includegraphics[width=\columnwidth]{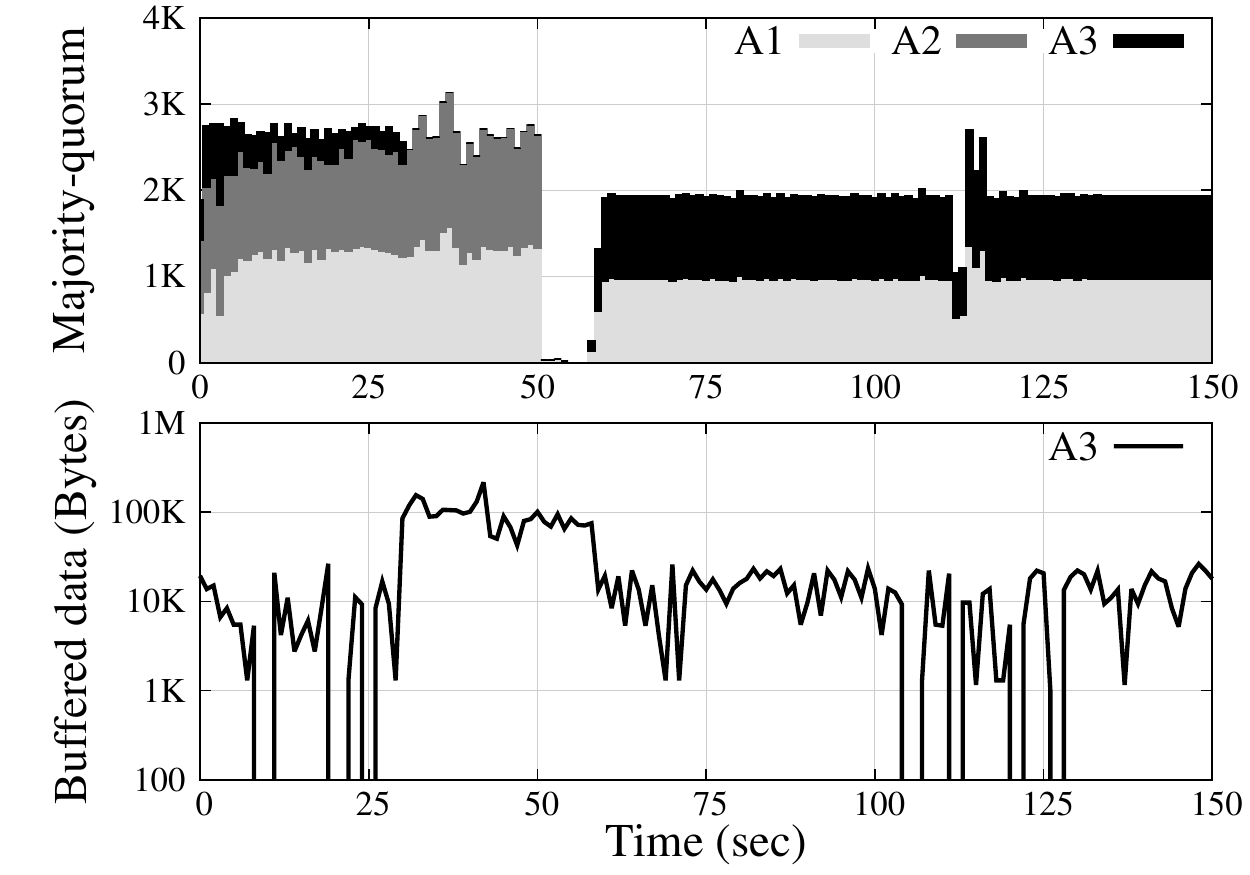}&
      \includegraphics[width=\columnwidth]{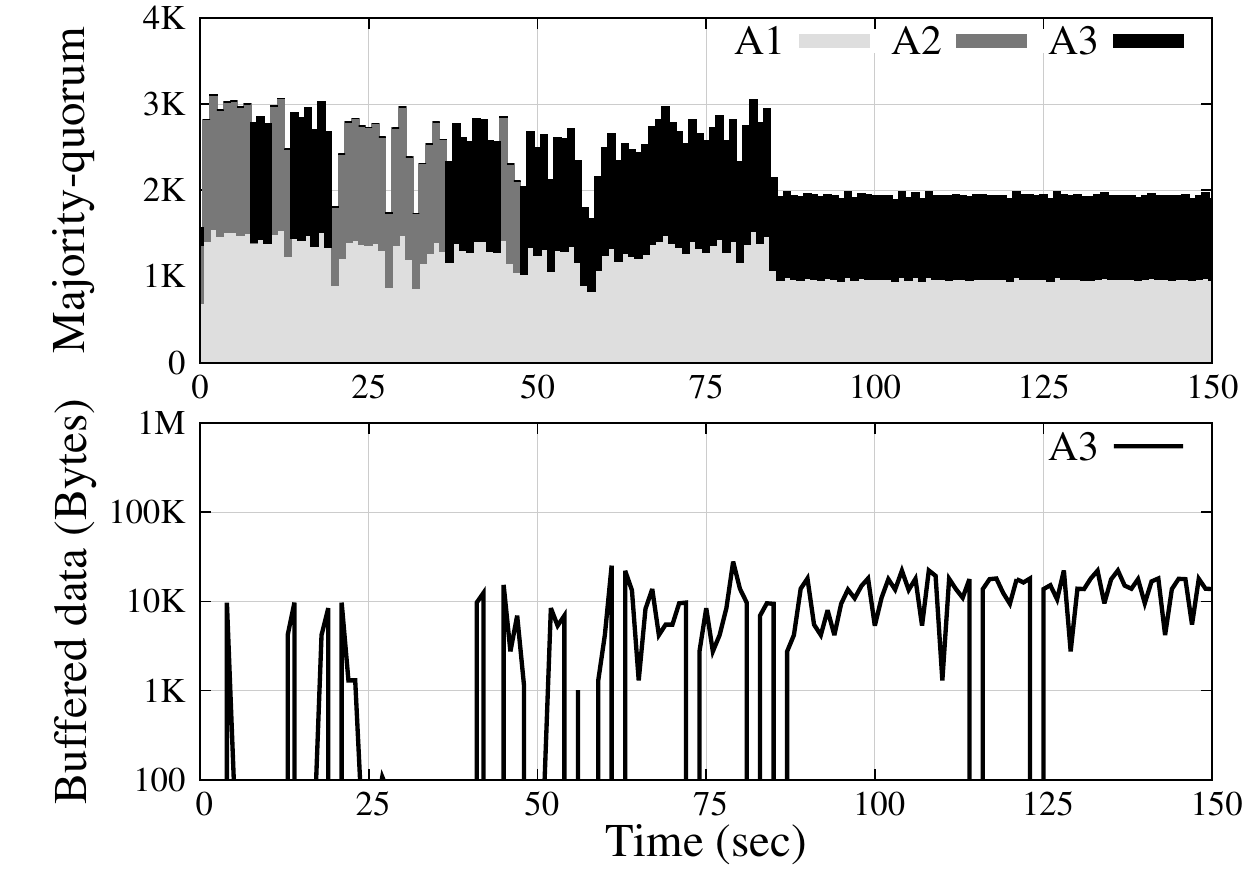} \\
    \end{tabular}
        \caption{Performance of \textbf{Libpaxos} (left graphs) and \textbf{Libpaxos$^+$} (right graphs) with 4 KByte requests at configuration (b) at 70\% of peak throughput (see Table~\ref{table:configs}); acceptor A2 is terminated after 50 seconds; majority-quorum for each acceptor measures the number of instances for which that acceptor is included in first majority-quorum;y-axis of the bottom-most graphs is in log scale.}
        
    \label{fig:lipaxos+:lan}
    \vspace{-4mm}
\end{figure*}

\begin{figure*}[ht]
    \begin{tabular}{c@{}c}
     \includegraphics[width=\columnwidth]{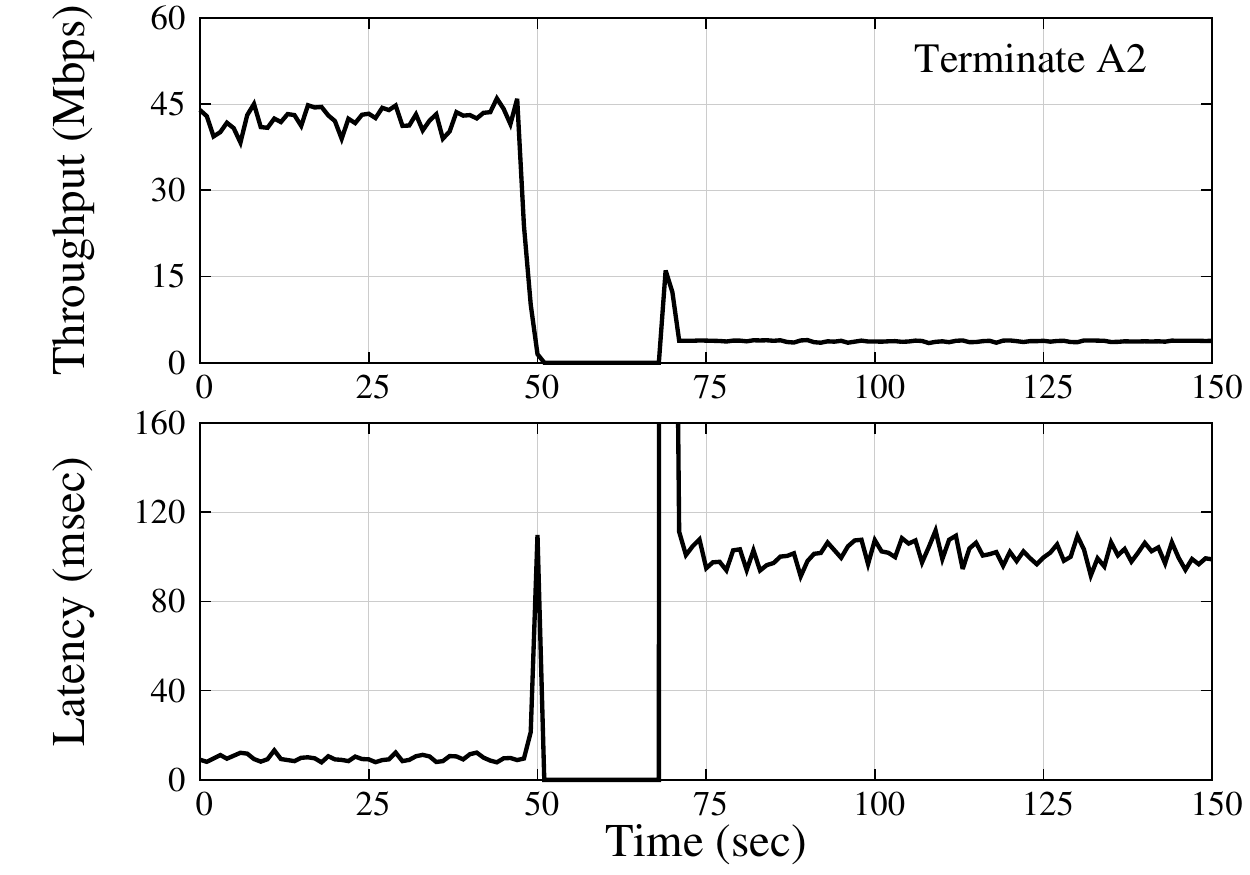}&
     \includegraphics[width=\columnwidth]{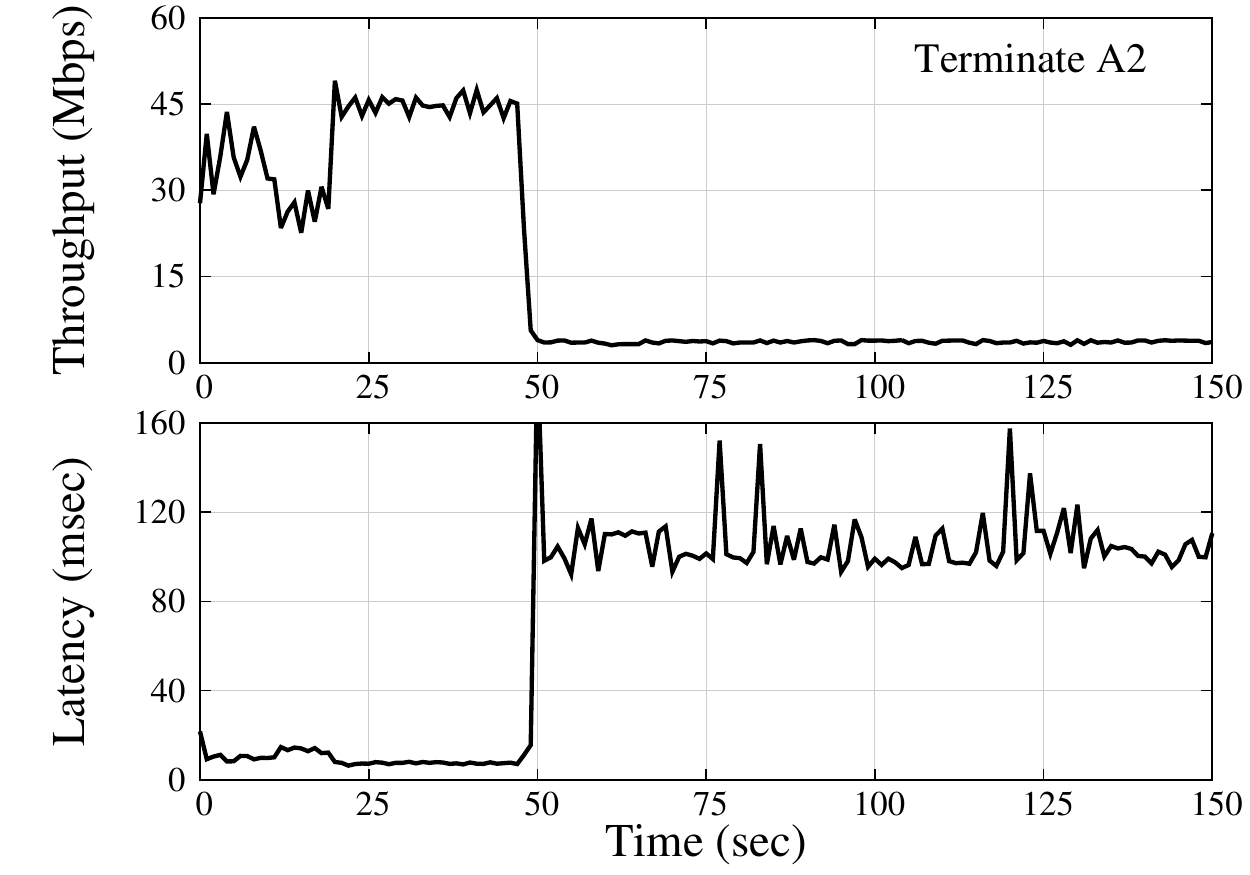}\\
    \end{tabular}
    \caption{Performance of \textbf{Libpaxos} (left graphs) and \textbf{Libpaxos$^+$} (right graphs) with 4 KByte requests at configuration (d) at 70\% of peak throughput (see Table~\ref{table:configs}); acceptor A2 is terminated after 50 seconds;}
    \label{fig:lipaxos+:wan}
    \vspace{-4mm}
\end{figure*}

\subsection{OpenReplica under failures} 
\label{subsec:openreplica}
The left-most graphs of figure~\ref{fig:openreplica} shows the performance of OpenReplica at configuration (c) with 100 KByte requests for a duration of 350 seconds at peak performance. In this execution, throughput varies between two values.
During intervals in which all the acceptors are responsive, throughput is higher. Throughput is lower when the leader needs to devote a fraction of its processing power to retransmit messages to the slow acceptor (see also Section~\ref{libs}). During this period, only the faster acceptors contribute to performance. We terminated acceptor A2 after 150 seconds of the execution when the throughput was at its lower value. As it is seen in the figure, performance suffers a small reduction at this point.  This is due to the fact that a majority-quorum is not immediately available, preventing the Paxos protocol from deciding new values until the quorum is restored. In the right-most graphs of Figure~\ref{fig:openreplica}, we executed OpenReplica in configuration (d) with 4 KByte requests under 70\% of peak performance. When acceptor A2 is terminated, after 150 seconds, throughput drops and latency increases since every majority quorum includes acceptors in different regions.
With both acceptors A1 and A2 operational, though, we can see that throughput oscillates. We also observed that while OpenReplica has bursty behavior under high load (i.e., peak performance in a LAN with medium and large values, and 70\% of peak performance in a WAN), it presents stable performance under moderate load.

\subsection{Ring Paxos under failures}
\label{subsec:ringpaxos}
Figure~\ref{fig:ringpaxos} shows the performance of Ring Paxos in configurations (b) (left-most graphs) and configuration (d) (right-most graphs) over a period of 350 seconds. In all the experiments after 150 seconds one of the acceptors (A2 or A3) is terminated. After an acceptor is terminated the performance drops to zero for a period of 2 to 3 seconds during which the ring is reconfigured. Left-most graphs show the performance of Ring~Paxos in configuration (b) with 100 KByte requests when the system is operating at its peak performance. When acceptor A3 is terminated, after 150 seconds, throughput increases. This is because Ring Paxos operates at the speed of the slowest acceptor (acceptor A3 in this experiment) and as soon as it leaves the ring, the protocol is no longer limited to its speed. This is also the reason why after terminating acceptor A2 the performance is not affected. Right-most graphs of Figure~\ref{fig:ringpaxos} show the performance of Ring~Paxos in configuration (d) with 4 KByte messages when the system is operating at 70\% of the peak performance. 
When acceptor A3, in the East coast is terminated (after 150 seconds), throughput increases and latency decreases. This is because Ring Paxos is no longer bound by the slow communication links of acceptor A3. Similarly to configuration (b) in configuration (d) after terminating acceptor A2 performance does not improve. 

\subsection{Libpaxos and Libpaxos$^+$under failures} 
\label{subsec:libpaxos}

In this section we consider the performance of Libpaxos and Libpaxos$^+$ in configurations (b) (Figure~\ref{fig:lipaxos+:lan}) and (d) (Figure~\ref{fig:lipaxos+:wan}) over a period of 150 seconds.
In these experiments, the request size is 4 KBytes and the system operates at 70\% of the peak throughput.
Acceptor A2 is terminated after 50 seconds of the execution. 
In configuration (b), A2 is a fast acceptor and in configuration (d) A2 is an acceptor located in the same region as A1. 
In both cases, the termination of A2 forces slow acceptor A3 to be part of a majority-quorum.
We report the following results in the graphs, from top to bottom: 
the delivery throughput in megabits per second, 
the latency as measured by the clients in milliseconds, 
the number of instances for which an acceptor's Phase~2B is included in that instance's first majority-quorum, 
and the amount of outgoing data buffered in the OS at acceptor A3. 
We recall that acceptors in Libpaxos forward values and Phase~2B messages to the learners and proposers. 


Before the termination of acceptor A2 in the experiments of Libpaxos (left-side graphs) in Figure~\ref{fig:lipaxos+:lan}, it is mostly acceptors A1 and A2 that participate in the majority-quorums. 
It can be seen in the third graph from the top that in the first 30 seconds of the execution, A3 participates in a few majority-quorums but then it becomes overwhelmed and data in its send buffers accumulates (see bottom graph).
%
When acceptor A2 is terminated (after 50 seconds), the delivery throughput drops to zero for a duration of 4 seconds, the time it takes slow acceptor A3 to process its backlog of previous instances. 
After acceptor A3 empties its buffers and can participate in the majority-quorum, the system becomes responsive. 
With acceptors A1 and A3, throughput is lower and latency is higher than in the beginning of the execution. 
In Libpaxos$^+$ (graphs on the right), after termination of A2, the shift to a new majority-quorum happens smoothly since A3 was never overwhelmed with requests. 
The proposer detects the slower acceptor A3 and spares it. 
Thus, when acceptor A2 is terminated, a majority-quorum is available immediately and the execution continues smoothly.
Notice that the sustainable throughput in Libpaxos$^+$ before and after the termination of A2 is similar to Libpaxos.

In Figure~\ref{fig:lipaxos+:wan} we investigate the behavior of Libpaxos and Libpaxos$^+$ in a wide-area deployment (configuration (d) in Table~\ref{table:configs}) with 4 KByte requests.
%
The behavior of Libpaxos after the termination of faster acceptor A2 in configuration (d) is similar to configuration (b), except that in a wide-area deployment it takes longer (18 seconds) for acceptor A3 to catch up with A1. 
After the execution resumes (at time 68), there is an important reduction in throughput and increase in latency when compared to the execution before the termination of A2. This is due to the large round-trip time between US-West and US-East regions. 
Similarly to Figure~\ref{fig:lipaxos+:lan}, the throughput of Libpaxos$^+$ does not drop to zero after the termination of acceptor A2.

In both Figures~\ref{fig:lipaxos+:lan} and~\ref{fig:lipaxos+:wan}, there is a peak in latency when normal operation resumes. 
This happens as the requests that the clients had sent immediately before the crash are only decided after acceptor A3 catches up. 

\section{Lessons Learnt}
\label{sec:lessons}

In this section we share the main lessons we have learnt from our experiments with the four open-source Paxos libraries.

Unlike Libpaxos and OpenReplica, S-Paxos and Ring Paxos allow clients to send their requests to any processes, which in turn disseminate the requests to other processes. 
The advantages of this scheme is that the protocol is not limited by the resources of only one process (e.g., network and CPU of the leader). 
Processes that receive client requests directly can batch and distribute them to other participants more efficient. 
The downside of this strategy is that in global deployments (e.g., configuration (d)) clients might select a process that is ``far away". 
One way to account for this limitation is to enhance the libraries to include policies for clients to select the process to which they transmit their requests considering the delay between the client and the process.

S-Paxos  and Ring Paxos operate at the speed of the slowest participant. 
Therefore, in heterogeneous configurations (e.g., configurations (b) and (c)) or when participants are distributed across multiple data centers (e.g., configuration (d)), Ring Paxos and S-Paxos are likely outperformed by the other libraries, which do not require all acceptors to be equally powerful as the leader or at the vicinity of the leader. 
If heterogeneity affects a majority of the participants, however, all the libraries will operate at the speed of the slowest member. 
Notice that leaving out a slow acceptor during failure-free scenarios has both advantages and disadvantages. 
The advantage is that in the absence of failures, the protocol operates at the speed of the fastest available majority. Libpaxos and OpenReploca benefit from this. 
The disadvantage is that during failures, the protocol might be stuck as it happens with Libpaxos but not with Ring Paxos and S-Paxos. 
(Ring Paxos has to reconfigure the ring, which introduces delays, but the reduction in performance is not due to the backlog of messages that gather at the slowest acceptor.) Ideally, a protocol would operate at the speed of the fastest members in failure-free scenarios and would not be penalized in case of failures, something we observed with Libpaxos$^+$. 
\pagebreak

While it may tempting for systems running on a tight budget to run Paxos with a majority of fast acceptors, adding a few slower ones purely as backups, our study shows that this strategy may seriously impact failover. 
Despite this, the implications of a heterogeneous quorum is often neglected in practice. 
We hope this paper will bring to the foreground the fact that performance differences in acceptors can have a significant impact on overall performance, especially when failures occur. 
We further suspect that the lessons we learnt apply to other quorum-based protocols, such as ABD~\cite{attiya1995sharing} and the initial Isis protocol~\cite{birman1990isis} since they all rely on a majority for progress. 


In summary, we can divide the protocols we surveyed in two categories: those that make progress by involving all participants (i.e., Ring Paxos and S-Paxos) and those protocols that make progress at the pace of a majority of participants (i.e., Libpaxos, Libpaxos$^+$, and OpenReplica).
The rationale of protocols in the first category is that performance can benefit from balancing the load among participants and exploiting the hardware resources of all participants.
A consequence of protocols in the second category is that slow participants do not hurt overall performance at peak load.
This design space has two important consequences.
Protocols in the first group are likely to suffer less with the failure of a participant, which is not the case with protocols in the second group: as the experiments show, Libpaxos's performance drops to zero and OpenReplica's performance becomes bursty after the failure of a fast participant.
The second consequence is related to the characteristics of the environment.
While homogeneous environments (i.e., nodes with comparable processing power and communication links with similar capacity) are more appropriate for protocols in the first group, protocols in the second group are more appropriate for heterogeneous environments.
Therefore, Ring Paxos and S-Paxos seem to fit better controlled environments such as private clouds and proprietary datacenters, while Libpaxos$^+$ seems the most appropriate for uncontrolled environments, such as Amazon's EC2, as it inherits the advantages of Libpaxos and OpenReplica while leaving out their disadvantages. 

\section{Conclusion}
\label{sec:final}
Paxos is one of the dominant protocols in building fault-tolerant systems and its performance has a significant impact on the overall efficiency of the systems built on top of it. Consequently, it is very important that Paxos implementations achieve steady performance and deal well with the load variations common in modern cloud computing settings. Our experiments reveal the large performance variations that out-of-the-box Paxos implementations may exhibit under even mild stress. We showed that without taking actions to stabilize the protocol, widely used Paxos libraries can exhibit sudden and rather long periods with no protocol decisions occurring at all. Our experiments also reveal surprising variability in the rate of decisions: some versions of Paxos are extremely bursty. Bursty throughput can cascade to create inefficient application-level performance. Finally, focusing on one Paxos implementation (Libpaxos), we showed how one can modify the protocol to preserve correctness and yet reduce the degree to which such problems arise.

\section*{Acknowledgment}
We wish to thank Daniele Sciascia (Libpaxos), Deniz Altinbuken (OpenReplica), and Martin Biely, Nuno Santos, and Zarko Milosevic (S-Paxos) for promptly answering our questions about the deployment and tuning of their libraries. We are especially grateful to Isaac Shef for showing us performance issues he encountered when working with a fifth Paxos implementation, and outlining his ideas for overcoming those problems. This work was funded, in part, by grants from the US NSF, DARPA, Swiss National Science and Zeno Karl Schindler Foundations.

\bibliographystyle{IEEEtran}
\bibliography{main,paxos-biblio}{}

\end{document}